\documentstyle[aps,aps10,epsfig,multicol,epsfig]{revtex}

\begin{document}

\title{Subcritical modulational instability and transition to chaos from periodicity}

\author{Hie-Tae Moon\cite{htmoon_mail}}

\address{Department of Physics, Korea Advanced Institute of Science
and Technology, Daejeon, 305-701, Republic of Korea}
\date{\today}
\maketitle

\begin{abstract}

This study shows that a modulationally destabilized monochromatic wave in a fluid system may undergo a subcritical bifurcation directly into chaos, when dissipation is weak enough. Analysis is made within the framework of the complex Ginzburg-Landau equations.  
\end{abstract}

\maketitle
\begin{multicols}{2}
This study concerns a subcritical transition to chaos directly from periodicity in a high dimensional system. We consider in particular a continuous physical system where a monochromatic wave is modulationally destabilized. The slow temporal and spatial modulation of the envelopes of a destabilized wave train, in a weakly nonlinear medium, is universally described by the Ginzburg-Landau equations(GL).\cite{newell} Within the theoretical framework of GL, this study reports that, when the strength of dissipation of the medium is weak enough, the modulated wave undergoes a subcritical transition directly into chaos. In the related earlier studies, it was generally shown that the modulationally destabilized wave would lead either to a supercritical transition into a 2-frequency motion or to a period-doubling bifurcation.\cite{doering,luce,rodriquez}

We start with the partial differential equation known as the Ginzburg-Landau equations;\cite{newell}

\begin{equation} 
\Psi _{t} = \epsilon \Psi + ( \epsilon + i ) \Psi _{xx} - ( \epsilon - i ) |\Psi |^{2} \Psi
\end{equation}

,where the real positive parameter $\epsilon$ measures the strength of dissipation of the system. When $\epsilon=0$, the system becomes dissipation-free and GL is reduced to the well known Nonlinear Schr{\"o}dinger Equation.\cite{yuen}

The GL is invariant under $x \rightarrow -x$, and therefore, for an even initial distribution under periodic boundary conditions with a period $-\pi/q < x <\pi/q$, the evolution can be expanded in terms of cosine functions as follows, 
\begin{equation}
\Psi(x,t) =  a_1 (t)+ a_2 (t) \cos qx + a_3 (t) \cos 2qx + \cdots 
\end{equation}
, where $a_i (t) $ are complex.

Substituting Eq.(2) into Eq.(1), we obtain, up to $ a_2 $\cite{ref1}, the following set of coupled equations;\cite{moon}
\begin{eqnarray}
 \dot{a}_1 &=&\epsilon a_1+(i- \epsilon)(|a_1|^2\, a_1 + 
 a_1|a_2|^2 +\frac{1}{2} a_1^* a_2^2 ), \nonumber \\
 \dot{a}_2 &=& \epsilon a_2- q^2 (i+ \epsilon) a_2 \nonumber \\ 
 &+& (i-\epsilon)({a_1}^{2}{a_2}^{*} + 2{|a_1|}^{2}a_2 +\frac{3}{4}{|a_2|}^{2}a_2), 
\end{eqnarray}
where $\dot{a_1}, \dot{a_2}$ denote respectively the time derivative of the 
complex variables $a_1(t), a_2(t)$. 

Eq.(3) now contains two parameters $\epsilon$ and $q$, which measure, respectively, the strength of dissipation of the system and the modulation wave number. When $\epsilon=0$, the system becomes conservative with a conservative quantity, $E = |a_1 |^2 + |a_2 |^2 $. This set of equations possesses a special symmetry property, namely, they are invariant under the symmetry $a_2 (t) \rightarrow - a_2 (t)$.  

Eq.(3) possesses the following limit cycle solution with frequency $\omega=1$;
\begin{equation}
a_{1}(t)=e^{it}.               
\end{equation}
A linear stability analysis shows that, for small values of $\epsilon$, the periodic orbit becomes unstable for the following perturbation
\begin{equation}
a_1(t=0)=1, a_2(t=0)=\delta, ~~\delta \ll 1,
\end{equation}
if
\begin{equation}
q \le q_{th} \approx \sqrt{2} (1-\epsilon^2). 
\end{equation}

In this study, we investigate the dynamics of the destabilized periodic orbit when dissipation is weak enough, say, when $\epsilon = 0.1$. For $\epsilon=0.1$, the periodic orbit begins to lose its stability at $q_{th} \approx 1.4$. As $q$ decreases from $ q_{th}$ to $q=q_c=1.236$, it is found that the frequency at $\omega =1 $ starts to shift gradually from $\omega = 1 $ to $\omega = \omega_c $, where the frequency is suddenly replaced by a broadband spectrum as shown in Fig. 1. Figure 2 gives the corresponding bifurcation diagram, where $K_i$, the $i$th maximum of the signal $|a_2(t)|$, is plotted for all $i=1,2,3,....$ on the vertical axis as a function of $q$ on the horizontal axis. The region denoted by ``$P_1$" or ``$ P_2$" in the bifurcation diagram corresponds to a periodic regime with a single frequency. The region ``$C$", on the other hand, is obtained from the motion exhibiting a broadband power spectrum. When the motion is in the region $C$, we further plotted $M_{i+1}$ as a function of $M_i$ for all $i=1, 2, 3,....$, where $M_i$ now denotes the $i$th maximum of the signal $|1-a_{1} (t)|$. The result is shown in Fig. 3, where we observe a non-invertible one-dimensional return map, indicating that the flow is now in a strange attractor. Incidently, we notice that this 1-d map has exactly the same shape as that of the Lorenz attractor\cite{lorenz}.

To understand the dynamics underlying the direct transition to chaos from periodicity, we now rewrite the solution as follows;

\begin{eqnarray}
a_1 (t)&=&1-Z(t) +i W(t), \nonumber \\
a_2 (t)&=&X(t) + iY(t)
\end{eqnarray}
, where $Z(t), W(t), X(t), Y(t) $ are all real.

We then attempt to make a geometrical analysis of the corresponding flows in the 4-d phase space, spanned by $X$, $Y$, $Z$, and $W$. Since one can not visualize trajectories in such a 4-dimensional space, we further reduce the dimension of the dynamics through the method of the Poincar\'{e}'s surface of section; Whenever a trajectory passes through the plane $W=0$ with a direction $\dot{W}>0$, we mark the other three coordinates (X, Y, Z) of the piercing point. The $n$th point ($X_n, Y_n, Z_n$) is then uniquely mapped into the next point $(X_{n+1}, Y_{n+1}, Z_{n+1})$, which now defines a discrete trajectory in a 3-dimensional space, which we refer to in this study as the Poincar\'{e} space.

The original limit cycle whose instability is under investigation is then represented in this Poincar\'{e} space as a fixed point located at the origin. Also the property that the system is symmetric under the reflection ${a_2} (t)  \rightarrow - {a_2 } (t) $  is now converted to the reflectional symmetry of the Poincar\'{e} space under $(X, Y, Z) \rightarrow (-X, -Y, Z)$. The fixed point at the origin in the Poincar\'{e} space becomes unstable at the threshold of linear instability $q = q_{th}$. Figure 4(a) shows that, when it becomes unstable, nearby trajectories are pushed away spirally into one of the pair of stable fixed points formed nearby. This type of bifurcation, referred to as a symmetry breaking bifurcation, results from the symmetry of flows under the reflection $(X, Y, Z) \rightarrow (-X, -Y, Z)$. This symmetry breaking bifurcation leads to the periodic regime denoted by $P_2$ in Fig. 2. As $q$ is further lowered to the critical value $q_c$, the trajectories are all of a sudden repelled from the new fixed point. Figure 4(b) displays the moment of transition, where it is seen that the neighboring trajectories of the new fixed point are spiraling out directly into a global double-wing structure. The noninvertible 1-d map displayed in Fig. 3, which has exactly the same shape as the 1-d map of the Lorenz attractor, comes from this structure. 
This double wing structure is therefore a strange attractor, which appears to have the same geometry of the Lorenz attractor. This provides a geometric explanation of the observed direct transition into chaos from periodicity. Incidently, readers are referred to Ref.\cite{malomed} for a derivation of the Lorenz model for the description of an infinitesimally weak amplitude modulation of a periodic continuous wave near the threshold of the modulational instability.

If it were a subcritical bifurcation, then the transition must also exhibit hysteresis\cite{berge}. By reversing the direction of $q$, we observe that the inverse transition takes place at $q= {q_c} '= 1.261$ instead of $q_c=1.236$. Such hysteresis is only possible if the pair of periodic orbits and the chaotic attractor coexist in the range $(q_c, {q_c}')$. Figure 5 actually depicts the presence of the invisible repulsive invariant torus (RIT) in the Poincar\'{e} space, which now separates the two basins; one for the global chaotic attractor and the other for the stable fixed points. This concludes that the transition is indeed a subcritical Hopf bifurcation.

Throughout the study, one may have noticed that there are strong geometric similarities between the discrete flows in the Poincar\'{e} space and the continuous flows of the actual Lorenz system\cite{lorenz}. We notice, however, one major difference. In the Lorenz system there exists the homoclinic connection, which is then followed by the transient chaos\cite{kaplan} that is to be converted into a chaotic attractor by a crisis\cite{grebogi}. We found no such homoclinic connection leading to a transient chaos in the present discrete flows of the Poincar\'{e} space. 

Although we have started with the continuous flow in a specific 4-d dynamical system, the conclusions drawn on the dynamics are based on the discrete flows in the reduced Poincar\'{e} space. The results therefore may bear a generic significance, and we present in Fig. 6 a schematic drawing of the bifurcation pattern leading to the subcritical transition to chaos from periodicity concluded in this study. Actually we observed the same subcritical transition to chaos in the continuous GL equations for the same initial condition. Figure 7 displays the bifurcation diagram found in the GL equations where the subcritical transition take place at $q=q_c =1.31$. 

To conclude, we have shown that a periodic orbit in a continuous system can undergo a direct transition to chaos when dissipation is weak enough. An analysis of the underlying geometry shows that the transition is accompanied by the emergence of the smooth manifolds of the Lorenz attractor in the 3-d Poincar\'{e} space. This study, through the idea of diffeomorphisms\cite{thompson}, may give an example which helps to understand the dynamic scope of the Lorenz attractor in a more complex, higher dimensional system.

\begin{acknowledgments}
 The author wishes to thank P. Coullet for helpful discussions. This work was supported by the Korea Science and Engineering Foundation(Grant No. R01-1999-00019-02002). 
\end{acknowledgments}


\pagebreak
\newpage
{ \bf Figures}
\psfull
\begin{figure}
\centering
\epsfig{figure=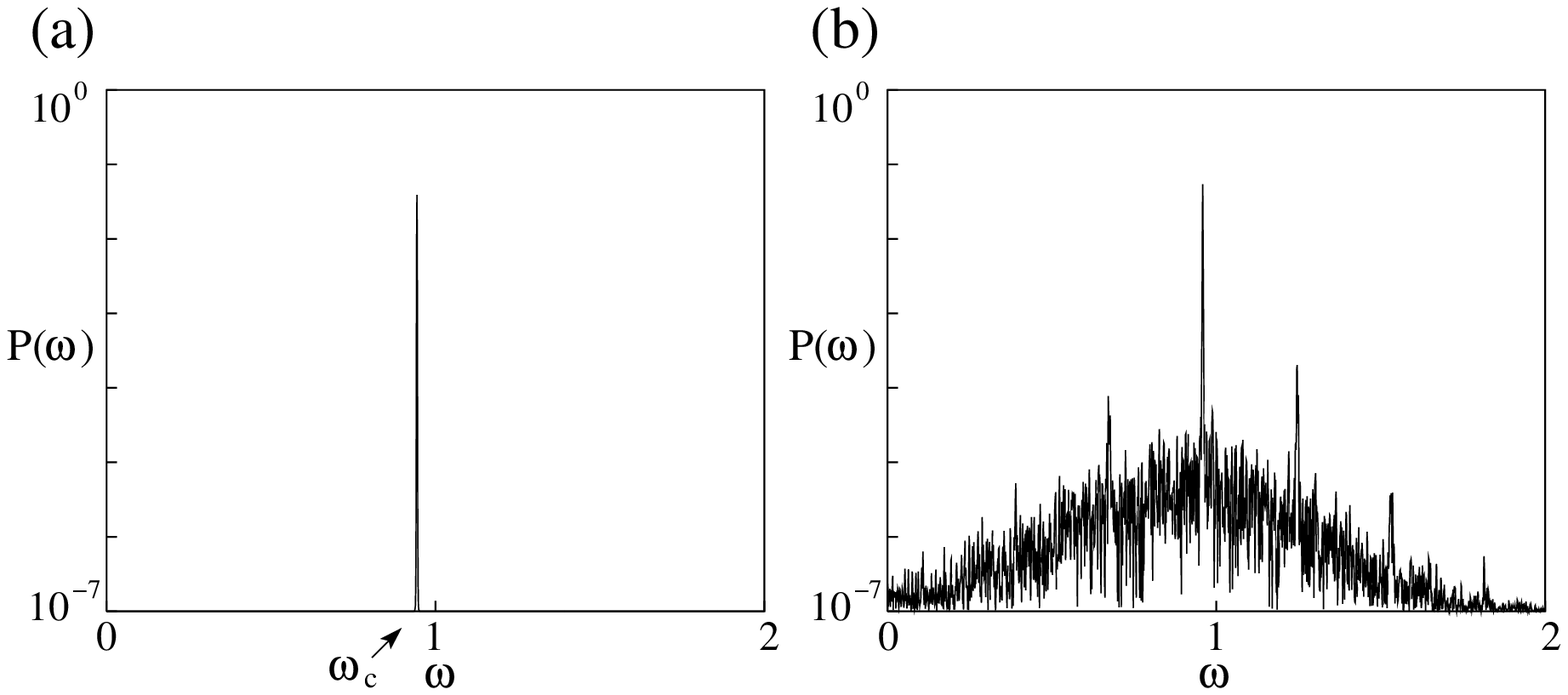, width= 9cm}
\caption{ Power spectra at (a) $q_{c}=1.236$, (b) $q =1.235 < q_c$.
}
\end{figure}
\vspace{0.5cm}

\psfull
\begin{figure}
\centering
\epsfig{figure=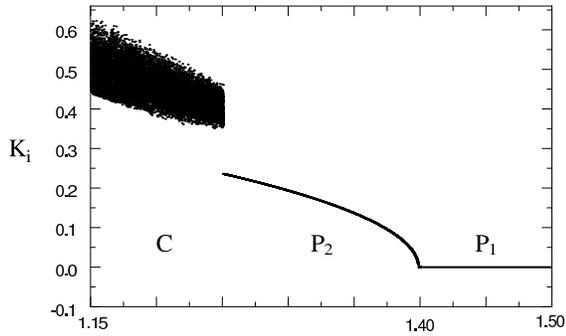, width= 7.5cm}
\caption{ 
Bifurcation diagram. $K_{i}$ denotes the $i$th maximum of $|a_{2} (t)|$. The ordinate displays $K_{i} $ for all $i=1, 2, 3, \cdots$. ``$P_1$" denotes the periodic state with frequency $\omega = 1$ before destabilization. ``$P_2$" denotes a periodic state for $q_{c} <q<q_{th} \approx 1.4$. At $q<q_c $, transition to chaos($``C"$) takes place. 
}
\end{figure}
\psfull
\begin{figure}
\centering
\epsfig{figure=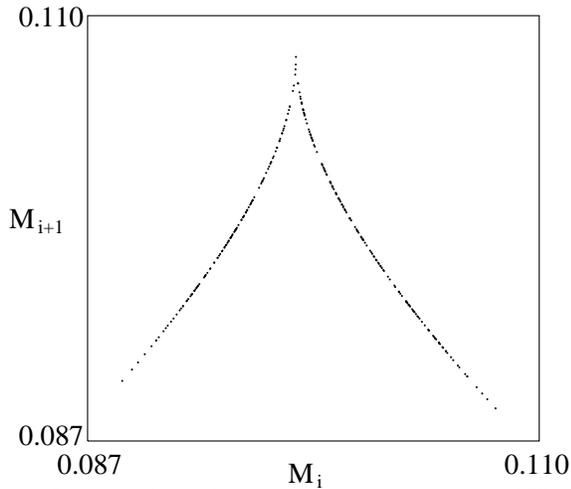, width= 7.5cm}
\caption{ 
One-dimensional return map for $q=1.235<q_c$. $M_{i}$ denotes the $i$th maximum of the signal $|1-a_{1}(t)|$. The abscissa shows $M_{i} $, while the ordinate displays $M_{n+1} $, for all $i=1, 2, 3, \cdots$.
}
\end{figure}
\psfull
\begin{figure}
\centering
\epsfig{figure=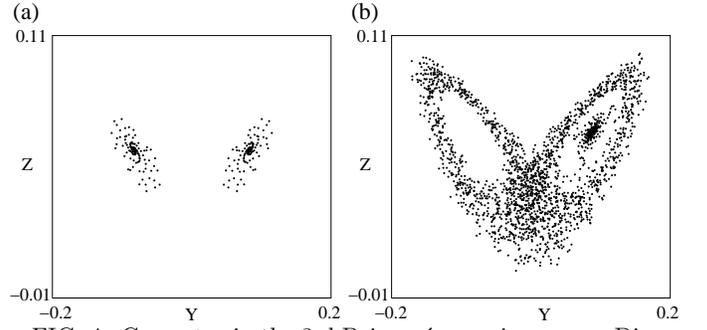, width= 9cm}
\caption{ 
Geometry in the 3-d Poincar\'{e} mapping space. Discrete trajectories are projected on the $Y-Z$ plane. (a) Pitchfork bifurcation of the fixed point at the origin. (b) Subcritical Hopf bifurcation. Trajectories starting near one of the fixed points are spiralled away into the outside global structure. The flow in this structure produces the 1-d return map shown in Fig. 3.
}
\end{figure}
\psfull
\begin{figure}
\centering
\epsfig{figure=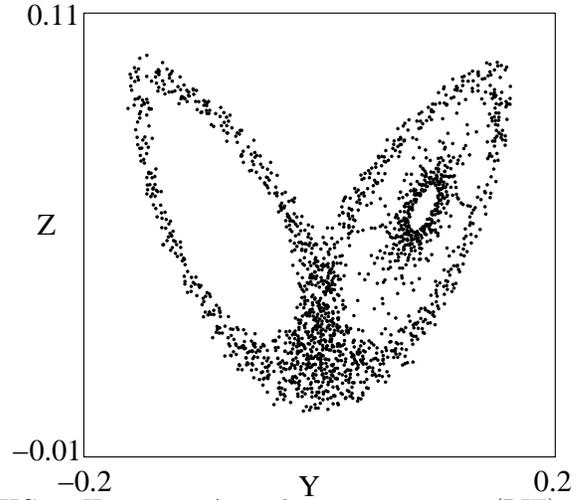, width= 7.5cm}
\caption{ 
Hysteresis. A repulsive invariant torus(RIT) separates two basins; one for the stable fixed point and the other for the outside chaotic attractor. Trajectories starting just outside of the RIT are spiraling out into the global chaotic attractor. 
}
\end{figure}
\pagebreak
\psfull
\begin{figure}
\centering
\epsfig{figure=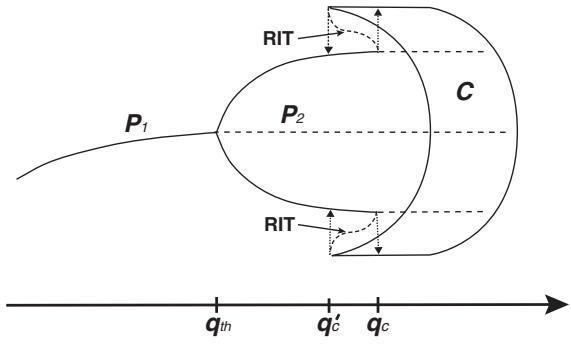, width= 7.5cm}
\caption{ 
Schematic drawing for the bifurcation pattern. $``P"$ denotes a periodic state and $``C"$ a chaotic state. The first bifurcation (at $q= q_{th} $) is a symmetry breaking bifurcation. The second bifurcation(at $q=q_c$) is a subcritical bifurcation into chaos($C$). Hysteresis is observed in the interval $[{q_c}' , q_c ]$. Following the usual convention, the dashed lines represents unstable states and the solid lines stable states. 
}
\end{figure}
\psfull
\begin{figure}
\centering
\epsfig{figure=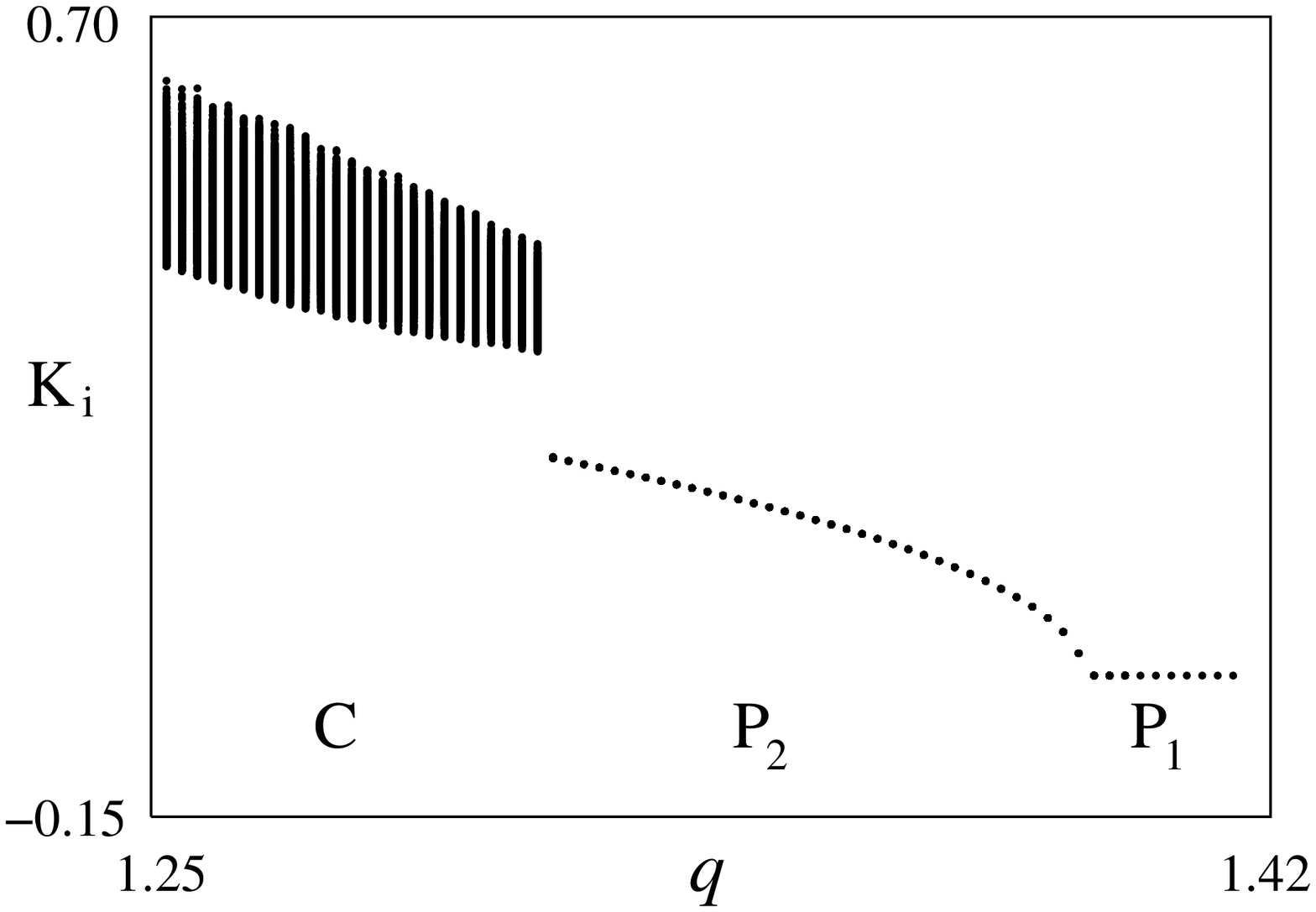, width= 7.5cm}
\caption{ 
Subcritical transition to chaos from periodicity in the Ginzburg-Landau equations for $\epsilon = 0.1$. See the captions of Fig. 2 for $ P_1, P_2$ and $C$. $q_c =1,3, q_{th}=\sqrt{2}$.
}
\end{figure}
\end{multicols}

\begin{thebibliography}{99}
\bibitem{ref1} The wavenumbers of the higher-order harmonics are all found to lie outside of the instability threshold $q_{th}$. Since the higher-order harmonics are all modulationlly stable, these are dropped. Later in this study, we will demonstrate that the dynamic implications concluded from the model are also shared by the GL equation.
\end{thebibliography}
\end{document}